\documentclass{article}

\usepackage{microtype}
\usepackage{graphicx}
\usepackage{subfigure}
\usepackage{booktabs} 
\usepackage{color}
\usepackage{xcolor}
\usepackage{amsmath,amssymb}

\usepackage{todonotes}

\usepackage[accepted]{icml2018}

\icmltitlerunning{Modeling Musical Onset Probabilities via Neural Distribution Learning}

\begin{document}

\twocolumn[
\icmltitle{Modeling Musical Onset Probabilities via Neural Distribution Learning}



\icmlsetsymbol{equal}{*}

\begin{icmlauthorlist}
\icmlauthor{Jaesung Huh}{equal,na}
\icmlauthor{Egil Martinsson}{equal,na}
\icmlauthor{Adrian Kim}{na}
\icmlauthor{Jung-Woo Ha}{na}
\end{icmlauthorlist}

\icmlaffiliation{na}{Clova AI Research, NAVER Corp., Seongnam-si, Gyunggi-do, South Korea}

\icmlcorrespondingauthor{Jung-Woo Ha}{jungwoo.ha@navercorp.com}

\icmlkeywords{onset detection, HazardNet}

\vskip 0.3in
]



\printAffiliationsAndNotice{\icmlEqualContribution} 

\begin{abstract}	
Musical onset detection can be formulated as a time-to-event (TTE) or time-since-event (TSE) prediction task by defining music as a sequence of onset events. 
Here we propose a novel method to model the probability of onsets by introducing a sequential density prediction model.
The proposed model estimates TTE \& TSE distributions from mel-spectrograms using convolutional neural networks (CNNs) as a density predictor.
We evaluate our model on the B{\"o}ck dataset showing comparable results to previous deep-learning models.
\end{abstract}

\section{Introduction}
\begin{figure}[!htb]
\includegraphics[width=1\linewidth]{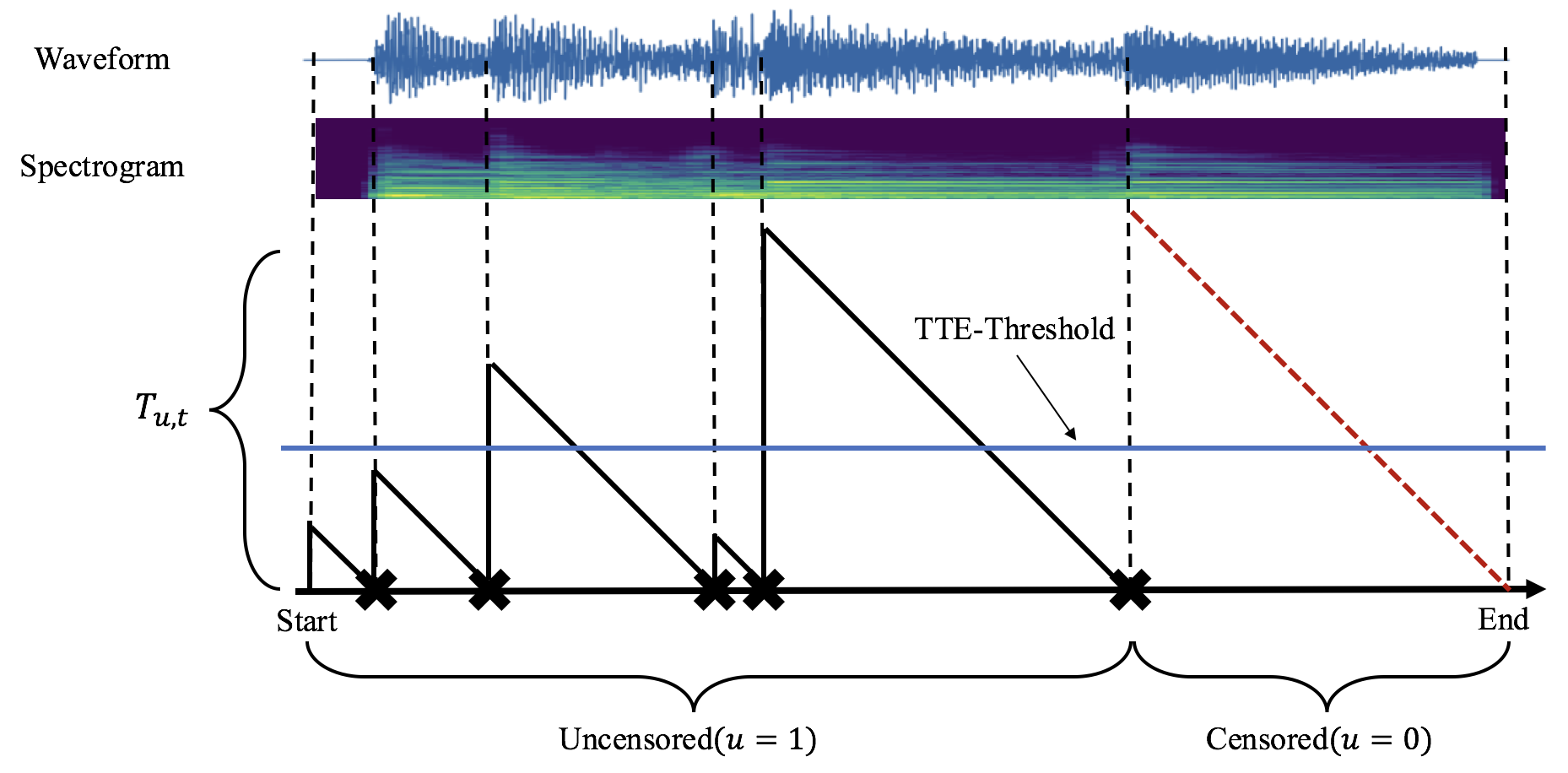}
\caption{Illustration of $T_{u,t}$ in Equation \ref{eq:loss} and censored data on an audio clip. An onset is notated as an X in the figure. We call the sequences after the last event \textit{censored}, because we do not know when the next onset will occur. One strength of our model is that the censored region can be used as training data by using a special loss function for censored data. TTE-threshold is a manually set upper-bound of time-to-event.}
\label{fig:tte}
\vskip -0.1in
\end{figure}

Musical onset detection is the task of finding the starting points of all relevant musical events in audio signals, which can be used in music-related applications~\cite{bock2012} such as automatic piano transcription~\cite{hawthorne2017}, rhythm game chart generation~\cite{donahue17}, and more.
Recently, many deep learning-based approaches have been proposed for onset detection such as RNNs \cite{eyben2010} and CNNs \cite{schluter2014}.
However, most approaches formulate onset prediction as a binary classification problem, which do not reflect the onset probability of frames adjacent to onset frames.
In this work, we formulate the onset detection problem as a combination of time-to-event (TTE) and time-since-event (TSE) prediction ~\cite{altman1998time, Martinsson2017}.
TTE is defined as the amount of time from the current time stamp to the next event (Figure \ref{fig:tte}), while TSE is the time elapsed since the most recent event.

\section{Proposed Model}
\begin{figure}[t]
\includegraphics[width=1\linewidth]{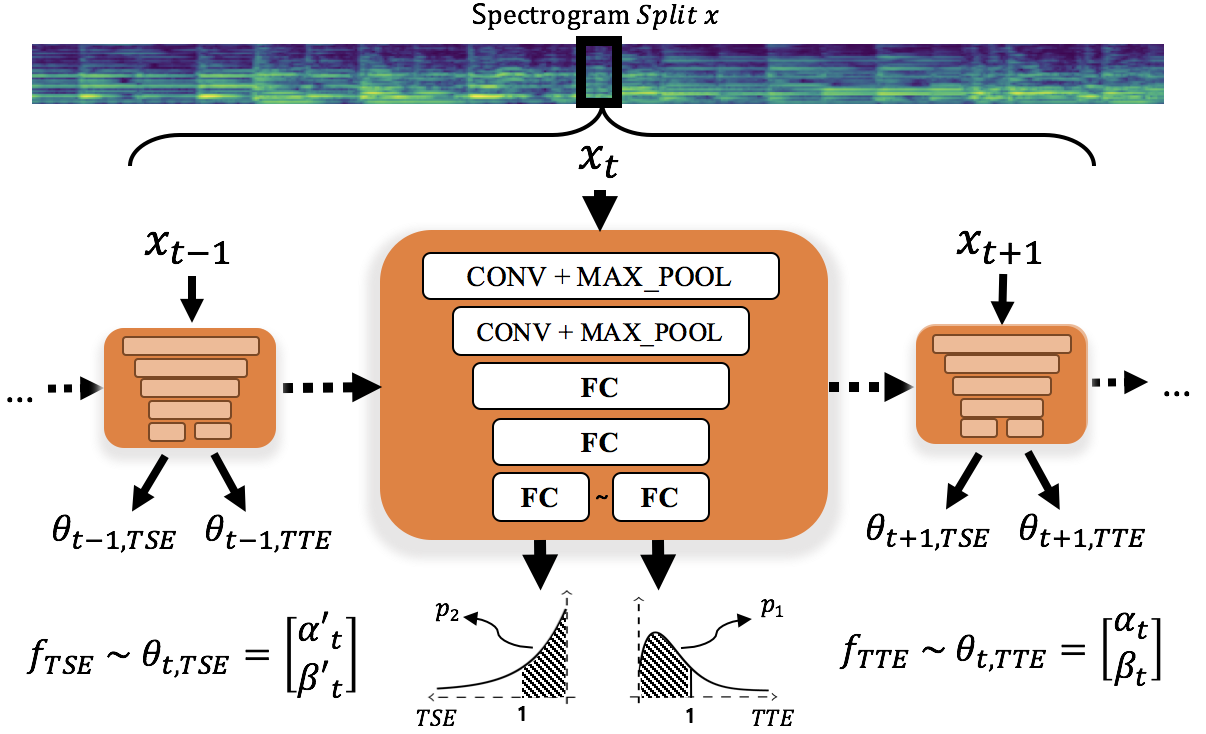}
\caption{Architecture of the proposed model. Our architecture is specified in Table \ref{table:model}. The predictor estimates the parameters of a target distribution to predict both TTE and TSE. }
\label{fig:architecture}
\end{figure}

Given an input feature sequence $\mathbf{x}$ transformed from sequence data such as a mel-spectrogram, an arbitrary neural network can be applied to fit the distribution of time-to-event as a density predictor. Here we use a convolutional neural network as the density predictor (Figure \ref{fig:architecture}).
At a given timestep $t$, the density predictor is fed with feature vector $\mathbf{x}_t$ to produce distribution parameters $\theta_t$ of TTE $tte_{t}\sim P_{\theta_t}$.
For a given timestep $t$ and its corresponding TTE distribution $P_{\theta_t}$, our model is optimized to minimize the negative log-likelihood for right censored data:
\begin{eqnarray}
    \label{eq:loss}L(T_{u,t}, P_{\theta_t}, u) = -\log(P_{\theta_t}(T_{u,t})^u S_{\theta_t}(T_{u,t})^{1-u}) \\
    \label{eq:survival}S_{\theta_t}(T)=P_{\theta_t}(tte_t > T)
\end{eqnarray}
TTE is called \textit{uncensored} (with indicator $u=1$) when we know the exact time to the next onset from time $t$ and \textit{censored} $(u=0)$ if we only observed a minimum bound, as illustrated in Figure \ref{fig:tte}. 
This means that when $u=1$ then $T_{u,t}$ is the TTE at the corresponding time stamp $t$ but time-to-end of sequence when $u=0$.
For uncensored TTE we maximize the likelihood of the next event happening at the corresponding time.
Otherwise, we optimize the likelihood of an onset occurring \textit{after} the end of sequence.
We have discrete TTE, so we model the discrete distribution using $P_{\theta_t}(T)$ as $F(T) - F(T-1)$ where $F$ is the cumulative distribution function. 
In addition to predicting TTE we predict time since last event by adding another dimension to the output layer, jointly predicting a distribution of both TTE and TSE.

\section{Experiments}
Experiments were performed on the B{\"o}ck dataset, which is well described in his work \cite{bock2012}.
Using Librosa \cite{mcfee2015}, we computed three log-magnitude 80 mel-scale spectrograms with different window sizes (23ms, 46ms, 93ms) and the same hop size 10ms to concatenate channel-wise.
The network input is 15 frame chunks with the decision frame positioned in the center.

We design our density predictor network based on previous work \cite{schluter2014}.
The main difference is that we have two separate output layers for predicting TTE and TSE, where each output nodes can be applied with diverse activation functions such as $softplus$ or $\gamma\times sigmoid$ to make sure parameters are positive and to enable training (i.e. $\gamma=5$) to be stable. For fair comparison on previous work, the baseline model is slightly modified to match the number of parameters. 
Details are specified in Table \ref{table:model}. Two $Dense 2$ layers are sharing parameters in our experiments.
In order to predict onsets from the estimated distribution, we compute the onset detection function (ODF).
At time $t$, we can formulate the ODF as the following, where $p_1 = P_{\theta_{t,TTE}}(tte_{t}\leq1)$ and $p_2 = P_{\theta_{t,TSE}}(tse_{t}\leq1)$.
\begin{eqnarray}
ODF(t) = 1 - (1-p_1)(1-p_2)
\end{eqnarray}
With computed ODF, we used a peak picking method based on equations below \cite{bock2012}.
The frame at $t$ is selected as onset if it satisfies the three conditions below: 
\begin{eqnarray}
\label{eqn:peakpick1}ODF(t) = max(ODF(t-t_1:t+t_2)) \\
\label{eqn:peakpick2}ODF(t) \geq mean(ODF(t-t_3:t+t_4)) + \delta \\
\label{eqn:peakpick3}t - t_{previous onset} > t_5 
\end{eqnarray}
In this work, we set $t_1$, $t_2$, $t_3$, $t_4$, and $t_5$ to 30, 30, 120, 10, and 0 ms for all experiments.
The evaluation is conducted by using mir-eval package \cite{raffel2014mir_eval} with 50ms tolerance. We calculated precision and recall by varying $\delta$ in (\ref{eqn:peakpick2}), and reported optimal F1-score for each fold, conducting 8-fold cross validation. We used the SGD optimizer and trained for 300 epochs for all experiments. We set the learning rate to 0.001 and momentum is linearly increased from 0.45 to 0.9 during 10 $\sim$ 20 epochs.
\begin{table}[t]
\caption{Structure of each model. For the proposed model, outputs are split in half to predict TTE and TSE after second BatchNorm. Dropout with $p=0.5$ is applied before each dense layer.}
\begin{center}
\begin{small}
\begin{tabular}{c|c|c}
\toprule
\multicolumn{2}{c|}{\textit{Proposed Model}} & \textit{Baseline}\\
\midrule
\multicolumn{3}{c}{Input $15 \times 80 \times 3$}\\
\multicolumn{3}{c}{BatchNorm}\\
\multicolumn{3}{c}{Conv $(7\times3)\times10$, ReLU}\\
\multicolumn{3}{c}{MaxPool 1x3}\\
\multicolumn{3}{c}{Conv $(3\times3)\times20$, ReLU}\\
\multicolumn{3}{c}{MaxPool 1x3}\\
\multicolumn{3}{c}{Dense 256, ReLU}\\
\multicolumn{3}{c}{Dense 20, TanH}\\
\multicolumn{3}{c}{BatchNorm}\\
\midrule
Dense 2 & Dense 2 & Dense 2, TanH\\
\multicolumn{2}{c|}{softplus, $\gamma \times$ sigmoid} & Dense 1, sigmoid\\
\bottomrule
\end{tabular}
\end{small}
\end{center}
\vskip -0.1in
\label{table:model}
\end{table}
\section{Results and Discussion~}
\begin{table}[t]
\caption{Experiment results as average F1-scores with standard deviations. Threshold indicates TTE\&TSE-Threshold. F1(S) indicates applying a hamming window with size 5 to the ODF.}
\vskip 0.15in
\begin{center}
\begin{small}
\begin{tabular}{lcccr}
\toprule
{} & Threshold & F1 & F1(S) \\
\midrule
  Baseline       & {} & 0.848$\pm$ 0.016 & 0.851$\pm$ 0.019\\
  LogLogistic                  & 5 & 0.878$\pm$ 0.015 & 0.874$\pm$ 0.017\\
  LogLogistic                  & 10 & \textbf{0.878$\pm$ 0.014} & \textbf{0.874$\pm$ 0.015}\\
  LogLogistic                  & 20 & 0.872$\pm$ 0.016 & 0.869$\pm$ 0.018\\
  Pareto                       & 5 & 0.843$\pm$ 0.023 & 0.840$\pm$ 0.024\\
  Pareto                       & 10 & 0.843$\pm$ 0.024 & 0.842$\pm$ 0.025\\
  Pareto                       & 20 & 0.837$\pm$ 0.025 & 0.837$\pm$ 0.026\\
\bottomrule
\end{tabular}
\end{small}
\end{center}
\vskip -0.1in
\label{table:result}
\end{table}

We tested with various thresholds (See Figure \ref{fig:tte}) on TTE as well as TSE, which reflects the characteristic of onset as a local event in an audio signal, with two-parameter distributions such as Pareto and LogLogistic.
Table \ref{table:result} shows that the LogLogistic distribution with Threshold 10 gives better results than the baseline, despite the architecture and number of parameters are almost the same.
This indicates that TTE prediction models can improve the accuracy compared to binary classification models.
Smoothing onset detection function with a hamming window does not help the performance in our case, but it helps in the baseline model.

For future work, further investigations on diverse distributions and model architecture should be conducted.
Our model can also be applied to more TTE prediction tasks in other domains such as medical and manufacturing fields.

\bibliography{main}
\bibliographystyle{icml2018}
\end{document}